# Prominence Perceptions as a Heuristic in Contexts of Low Information

Esteban Villa-Turek

*Abstract* - This study explores the concept of prominence as a candidate trait, understood as the perceived worthiness of attention candidates elicit from regular citizens in the context of low information elections. It proposes two dimensions of candidate prominence, political and public, operationalized as having held high visibility roles within the party and having social influence through social media presence. Employing a conjoint analysis experimental design, the study tests whether political and public prominence serve as heuristic mechanisms in low-information electoral settings by estimating conditional effects on respondents' self-assessed interest in politics, educational level and self-assessed ideological placement. The results contribute experimental evidence to support the hypothesis of differential heuristic choices by voters based on varying levels of perceived public and political prominence, conditional on voters' characteristics.

# Introduction

Valence theory was initially introduced by Donald Stokes to account for all elements of electoral choice found empirircally in the real world by having stable policy-oriented (*strong ideological focus*) choices and diffuse valence-oriented (*weak ideological focus*) choices (Stokes, 1963), which incorporates features of real life by embracing the fact that usually what drives voters' choices rests on potential performance or viability perceptions of the party or cadidate to address and manage the most important issues according to the voters (Sanders et al., 2011). Such perceptions include perceptions of successful past performance, competence, or viability and belong to the same cognitive factors that Stokes referred to, also called "fast and frugal heuristics" (Goldstein & Gigerenzer, 2002) in the psychology and decision making literature.

These heuristics act as mental shortcuts taking advantage of the scarce information that may be available in most real-world contexts. For instance, voters may use the perceived competence of a party's leader as a mental shortcut to establish how likely the party will be to manage difficult situations in which all or nearly all citizens share similar goals (e.g., defending the country from foreign attack, or avoiding recessions. A similar role is played by the attitudes towards the perceived ability of a party or candidate to navigate the most pressing needs of the society according the the majority of voters. Party identification may be especially important because it can serve as a heuristic when employed as summary of issue stands and future policy actions, and/or as a sort of perceived competence track record bewteen parties or candiates (Sanders et al., 2011).

This is the case particularly in contexts of low-information elections:

> "(…) these elections are the rule, not the exception, in American Politics. Yes. Citizens are frequently asked to weigh in on races where the most effective piece of information—partisanship—is unavailable. These races include prominent positions, such as mayor, but also less visible positions, such as court clerk, public defender, school board members, city council members, and local authority positions, such as port commissioner and fire commissioner.

> Additionally, political primaries require citizens to adjudicate between candidates who are indistinguishable on party lines." (Kam & Zechmeister, 2013, p. 971-972)

The reliance on heuristics when voting, however, does not mean that valence driven electoral choices are less rational, much less irrational. In fact, the definition of rationality that this study employs aims at getting to a more nuanced understanding of rationality in uncertain contexts, where information is scarce and costly to obtain. This characterization of rationality finds grounding on the notion of low-information rationality (Simon, 1955) or, better yet, of models of ecological rationality (Goldstein & Gigerenzer, 2002), according to which the heuristic-driven choices may be rational because the heuristics are positively correlated with the outcome being predicted (Goldstein & Gigerenzer, 2002).

Given that recognition is valid due the informational structure in which it is embedded, this study proposes a subjective perspective according to which the informational structure varies with each voter's particularities. To do so, it focuses on one aspect of recognition that has not been studied thus far regarding its implications for electoral races: a candidate's promincence. The study builds on Simon Munzert's (2018) novel approach to calculate politicians' importance using Wikipedia data, but only regarding the conceptualization he offers of political importance: "Political importance is considered to be the combination of *prominence*, subsuming characteristics that contribute to the popular perception of politicians, as well as *influence*, describing how well politicians are connected among their peers and their footprint in the political arena." (Munzert, 2018, p. 27). In this sense, prominence is a latent component of overall importance and can be thought of as the worthiness of general attention that a candidate elicits from the public (Munzert, 2018), which is crucial especially if it is clear that the a person's attention is inherently limited. The latter plus the real-world challenge of low information elections, where every possible cue of candidate viability can be a decisive factor, warrants the study of a candidate's prominence as an objective recognition object, building on previous research on name recognition (Kam & Zechmeister, 2013).

Therefore, the study will ask the following research questions:

    RQ1: do perceived levels of prominence drive voter choices?

    RQ2: is there any difference between perceptions of public prominence and political prominence?

    RQ3: is the difference attributable to informational structures that vary depending on voter characterstics like political interest or ideological position?

# Theory

Anthony Downs' *An Economic Theory of Democracy* (Downs, 1957) introduced an economic approach to understand democratic political processes through rational-choice models. His theory is based on a series of important assumptions that include that all decisions are made centrally in the government, government has only two choices at a time, the choices are independent of each other, the framework is that of a two-party system, parties know what the preferences of all voters are, and voters know all possible governmental and party choices and their consequences (Downs, 1957, p. 54). The last assumption is key, as it implies perfect information based on which voters can calculate which party (or candidate) to support to maximize their preferences (Downs, 1957). However, in Chapter 5 Downs introduces the notion of uncertainty as "any lack of sure knowledge about the course of past, present, future, or hypothetical events" (Downs, 1957, p. 77). Downs recognizes that uncertainty is important, because it affects the confidence with which agents in the models make choices (Downs, 1957). In fact, Chapter 6 opens the door to a more plausible rational decision-making process that is not rational merely on the account of being based on perfect information but rather taken its context of low and costly information. Downs argues that uncertainty makes voters fall into different types relative to the confidence they have in their electoral choices (Downs, 1957). This chapter therefore introduces an important alternative of rational decision making taking into account different levels of access to information on the voter's side (Simon, 1955) and that later developed as valence theory, introduced by Donald Stokes to account for all elements of electoral choice found empirircally in real-world contexts, as noted above (Stokes, 1963).

This warranted the study of heuristics in the political realm based on psychological research about decision-making in low-information contexts and the apt use of heuristics to navigate the difficult access to infurmation (Goldstein & Gigerenzer, 2002; Simon, 1955). Such is the case with the most basic and general heuristic, the recognition heuristic:

> "(…) heuristics (…)are (a) ecologically rational (i.e., they exploit structures of information in the environment), (b) founded in evolved psychological capacities such as memory and the perceptual system, (c) fast, frugal, and simple enough to operate effectively when time, knowledge, and compu- tational might are limited, (d) precise enough to be modeled computationally, and (e) powerful enough to model both good and poor reasoning. We introduce this program of fast and frugal heuristics here with perhaps the simplest of all heuristics: the recognition heuristic." (Goldstein & Gigerenzer, 2002, p. 75)

These recognition-enabled inferences are ecologically valid because they are embedded in a particular informational structure (like low-information elections), where a lack of relevant information is systematically distributed and therefore strongly correlated, in either direction, with the criterion being predicted (Goldstein & Gigerenzer, 2002). Since the criterion being predicted is unknown, for example the endoment of a university or the size of a city in foreign country, the validity or strength of the recognition heuristic can be explained as a tripartite process in which a mediator is used as a source of information to infer the unknown criterion (Goldstein & Gigerenzer, 2002). Thus, the ecological validity is the relationship between the mediator and the unkown criterion and the surrogate validity is the relationship between the mediator and the mind of who infers (Goldstein & Gigerenzer, 2002). To clarify the latter and to express how good an inferential mechanism the recognition heuristic is, Goldstein & Gigerenzer (2002) conduct a recognition test in which all mentions of all German cities with more than 100,000 inhabitants appeared in the Chicago Tribuen from 1985 until 1987 and compare it with a similar study performed in Austria with mentions of the largest American cities in Die Zeit. The results can be found in Figure 1.

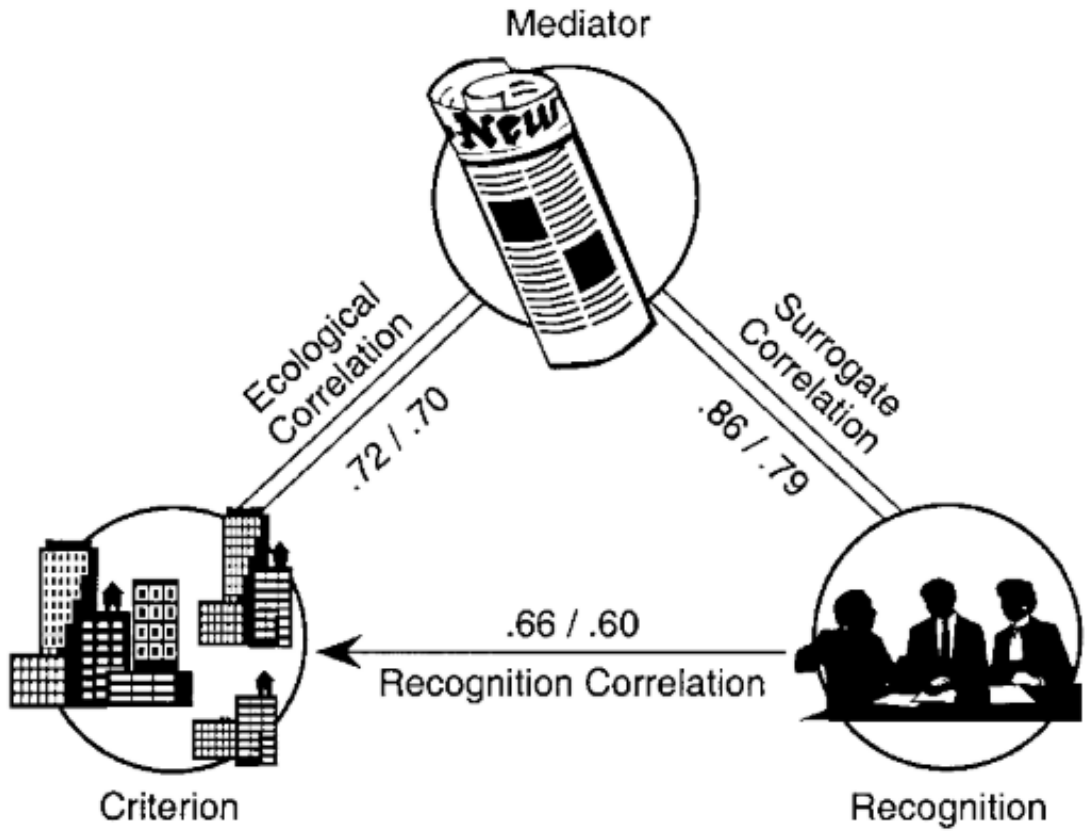

**Figure 1** – "Ecological correlation, surrogate correlation, and recognition correlation. The first value is for American cities and the German news- paper Die Zeit as mediator, and the second value is for German cities and the Chicago Tribune as mediator. Note that the recognition validity is expressed, for comparability, as a correlation (between the number of people who recognize the name of a city and its population)." (Goldstein & Gigerenzer, 2002, p. 86)

In Figure 1, the first correlation value for each relationship represents the results of a test for American cities mentioned in a German news outlet, and the second represents the results of the test for German cities mentioned in an American news outlet. As can be seen, the ecological and surrogate correlations are stronger than the recognition correlation, which is the ultimate inferential task of interest. This seems natural as the recognition is mediated by the number of times the cities appeared in the newspaper, meaning that most of the participants recognized accurately the largest cities in the newspapers, the surrogate correlation, as it is is the only direct cognitive contact that the respondents had with any information regarding the cities. As the relationship grows more indirect, i.e. where people are not directly recognizing anything, the ecological correlation decreases, and although the final recognition correlation is the lowest of them all, partly because of the uncertainty surrounding it, the correlation coefficient is nevertheless noteworthy.

In the realm of political science, valence attributes seem to drive voter choice more often than not. Empirical evidence has found support for the coexistence of both valence and spatial cognitive processes in voting, with the caveat that the direct effects of valence considerations on electoral choice tend to be stronger than their non-valence counterparts (Sanders et al., 2011). Such is the case, for example, of the recognition heuristic, which has been rightfully identified as a crucial driver of choice in low-information voting contexts (Kam & Zechmeister, 2013; Panagopoulos & Green, 2008). Furthermore, forecasts based on simple recognition heuristics have been found to be accurate in contexts of multiparty elections regarding smaller, more obscure political parties (Gaissmaier & Marewski, 2011), a finding that may well apply to contexts of low information elections in bipartisan systems regarding obscure candidates or nominees, like the United States. For instance, where party affiliation determines a significant part of electoral outcomes, candidates often find themselves juggling a wide array of attributes, both policy and non-policy related, in order to get elected (or reelected), despite belonging to either one of the principal political parties (Ansolabehere et al., 2001). Another example of a candidate's valence attribute driving voter support can be seen in gender among Democratic candidates, resulting in female Democratic candidates eliciting more support among liberal voters in low-information contexts (Mcdermott, 1997). More recently, studies on the effect of personalization strategies of candidates using social media have found that such strategies do elicit a higher awareness of a candidate conditional on voter characteristics, therefore increasing the likelihood of voters' heuristics on candidates' personal traits (McGregor, 2017). Regarding age, a recent study found significant effects of a candidate's age as heuristics when the candidate shares the same party affiliation and is closer in age to the voter, but more importantly the heuristic's effects vary with contextual and voter-specific characteristics (Webster & Pierce, 2019). In a comparative setting, studies in Brazilian low-information elections seem to support the idea of voters using heuristics related to the personal qualification of the candidates, finding general support for candidates with the title of "doctor", but differential support (or lack thereof) when the candidate's title is "pastor" (Boas, 2014).

In studying a specific form of the recognition heuristic, name recognition, Kam & Zechmeister (2013) argue that there are two causal pathways through which name recognition influences voters' candidate support. The first causal pathway is the direct one and draws from the psychology and consumer marketing literature on mere exposure. In short, the direct causal pathway suggests that voters will tend to favor the candidates to which they have been exposed the most (Kam & Zechmeister, 2013). The indirect causal pathway, on the other hand, draws from the literature on decision science and posits that the recognition of a candidate allows voters to make inductive inferences about the candidate (Kam & Zechmeister, 2013).

In the case of electoral processes where the recognition heuristic plays a role, the correlation between candidate recognition and the ecologically valid prediction is positive and the inferences made have been found to relate to the candidate's viability, rather than to her traits or experience (Kam & Zechmeister, 2013). In fact name recognition seems to drive electoral choices in low information elections because recognition tends to translate into higher candidate support (Kam & Zechmeister,

2013). Nevertheless, the more interesting finding relates to the plausible interplay between name recognition and other factors like incumbency, appearance, partisanship, ethnicity, prominence, etc. This is important, because, differently to what research on the recognition heuristic has contributed, electoral choices are *matters of taste or judgment* and not strictly probabilistic inferences. This allows for the recognition heuristic to acquire a more compensatory nature in the presence of other cues that could potentially be more germane to the electoral choice at hand, thus diminishing the indirect effect of recognition on voter choice but not eliminating it (Kam & Zechmeister, 2013). This possibility opens the door for a more nuanced muti-dimensional apporach to study voter heuristics in which several cues about competing candidates are evaluated simultaneously, thus resembling the comlplex choices that voters actually face.

Given the tripartite structure of the recognition heuristic, the study proposes a differential ecological validity conditioonal on voter characteristics. For instance, if a voter is very interested in politics, she will access a different informational structure with a distinct ecological validity than a voter who is not, which also implies that particular cues will have a differentiable effect on overall candidate support. To assess how voter characteristics may determine which cues have a larger effect in electoral choices, the study introduces a distinction of prominence as a component of a candidate's overall importance. Thus, one variant of prominence will be strictly political, aimed at signaling political expertise or competency by means of indicating if the candidate has held a more obscure public office before, like city council member (yielding *low* political prominence) or governor (*high* political prominence). The other variant will be more general and intended to signal prominence to the general public by indicating how many Twitter followers a candidate has.

# Methodology

The study employs a conjoint analysis methodology hereinafter, following the description and motivations presented in Hainmueller at al. (2014). Conjoint analysis saw its birth in the early 1960s with an application of mathematical psychology that allowed for the measurement of simultaneous combinations of quantities from the same kind. It was later revisited as a way to measure consumer preferences and decision-making in the context of complex and multidimensional choice scenarios (Hainmueller et al., 2014). It has been widely used by marketing specialists to research consumption behavior, product development and preference formation, with diverse variations having been sparsely used in sociology as well, under the name of 'vignettes' or 'factorial surveys' (Hainmueller et al., 2014). The inclusion of conjoint analysis into political science serves the need for a better way to infer causality between experimental manipulations and observed phenomena (the composite treatment effects), as a tool to clearly identify causal effects of individual elements of any kind of treatment in a survey experiment. This task is normally hard to achieve by the traditional survey design, which only allows the researcher to estimate causal effects as a whole, but not the individual and specific effects from single elements of the experimental manipulation (Hainmueller et al., 2014). The latter takes on particular importance when planning an experiment that will explore the effect of single changes on multidimensional choices, such as when the research simulates a hard choice setting for respondents to choose the hypothetical immigrant's profile that will be granted entry into the country or to select the hypothetical political candidate's profile for whom she would vote for (Hainmueller et al., 2014; Hainmueller & Hopkins, 2014). Other examples of conjoint analysis designs can be seen in Franchino & Zucchini (2015) with a similar political candidate experiment ran with undergraduate students in Milan; in Hainmueller et al. (2015) with a research on the Swiss population's attitude towards immigrants, which was later compared to a natural experiment caused by a referendum on the same subject with remarkably good results; or in Carnes and Lupu ( 2016), where in a comparative study, North American, British and Argentinian respondents were asked to choose between candidates, with the aim of measuring whether they disliked those who were working-class candidates (which they did not), to name just a few.

Particularly of interest is the variation of the conjoint analysis technique that randomizes the display of the different treatments of interest, allowing for a decomposition of the composite treatment effects. By means of the identification of a causal quantity of interest, the average marginal component effect (the AMCE hereinafter), and by making a series of assumptions that necessary hold because of the experimental design itself, the AMCE can and will be nonparametrically identified from the conjoint data collected in the survey experiment (Hainmueller et al., 2014). It is noteworthy that the nonparametric nature of the estimation of the AMCE allows for the researcher to avoid resorting to assumptions of functional form, simplifying the statistical approach greatly, mainly because no assumptions of behavioral models of respondents need to be made in order to fit the observed data efficiently. Therefore, the conjoint analysis method does not need any assumption about the behavioral model under which respondents formed preferences and made their choices to allow for an efficient and, above all, unbiased AMCE identification (Hainmueller et al., 2014).

A conjoint analysis has several other advantages as well, beyond the practical and convenient property of causal effect identification of individual components, as referred to above. As Hainmueller et al. (2014) explain, conjoint analysis provides, first, a sense of realism when presenting complex and multidimensional choice settings to respondents like the ones they would encounter in the real world. Second, it allows for a simple, cost-effective way of testing multiple hypotheses within the same experimental design. Third, and linked to the latter, it allows researchers to evaluate whether different theories have or lack explanatory power, by means of a single experiment with a single behavioral outcome that estimates the effect of multiple treatment elements at once. Fourth, the risk of social desirability bias in the respondents' stated choice preferences are significantly reduced since the respondents can justify their choices by means of any of the numerous other treatment elements simultaneously at play. And fifth, conjoint analysis can exploit its marketing forecasting potential for practical problems, such as policy design, if, for instance, it was used to predict the most popular policy elements combination in an upcoming reform (Hainmueller et al., 2014).

By design, there are several assumptions to be made to estimate the AMCE correctly, all of which are held by the design of the conjoint analysis experiment itself, or by testing with observed data (Hainmueller et al., 2014). Below are the most important and most pertinent ones for this research design, as explained by Hainmueller at al. (2014):

1. We first assume no carryover effects and stability, meaning that the current choice made by the respondent is not influenced by the last choice made by her, given the treatment effects presented in that choice task. She will always choose based on the same treatment when it appears, no matter what other treatments preceded the current task. It also means that potential outcomes remain stable across all possible choice tasks.

2. The second assumption is no profile-order effects. It allows researchers to ignore the order, if any, in which the different attributes are presented to the respondents, allowing for the former to simply pool information of interest across profiles for estimation purposes. This assumption helps boost the efficiency of the conjoint analysis.

3. The third assumption is randomization of profiles and implies that the outcomes are statistically independent of the profiles. By design, the conjoint analysis should present randomized attributes as profiles to the respondents and therefore the choices they make will not be systematically related to the profiles they see. Moreover, each level of each attribute must have a non-zero probability of being randomly presented in a profile (unless there is theoretical reason to define prohibited pairs of attributes that would not make sense in real life).

Based on the latter assumptions the design allows for the estimation of the individual effect of any given treatment component, or AMCE. The goal is to understand how any given treatment component affects the probability of a profile being chosen while having under consideration that such individual effect may be -and usually is- different depending on the other attributes of the profile, which allows for the estimation of the marginal effect of the treatment attribute "averaged over the joint distribution of the remaining attributes" (Hainmueller et al., 2014, p. 10). We can also estimate interaction effects between treatment components, for instance income and public prominence of the candidates. This interaction effect is the average component interaction effect, ACIE, and can operate, as mentioned above, as the interaction of two treatment components, where the ACIE of the two treatment components of interest is the difference in percentage point estimates in average marginal component effects of the income level between a candidate with a high level of public prominence and a candidate with a low level of public prominence. Furthermore, the interaction effect can be estimated between any given treatment component and a characteristic of the respondent, like age or political ideology (Hainmueller et al., 2014).

Finally, as estimation strategies it is possible to perform a simple difference in means or a linear regression (Hainmueller et al., 2014). The study assumes completely independent randomization of treatment components, that is, the candidate profiles can take on any combination of possible attributes, without any restriction or prohibited pairs. This way, it is possible to estimate the AMCE as the difference in means between the number of profiles where the treatment component occurred and the number of profiles where it did not occur or by fitting a linear regression "of the observed choice outcomes on the (...) dummy variables for the attribute of interest and looking at the estimated coefficient for the treatment level." (Hainmueller et al., 2014, p. 16). Furthermore, and very conveniently, we can also estimate the AMCE of all treatment components by simply regressing the outcome variable on the sets of dummy variables for every attribute level (excluding the baselines) and thus the AMCE can be interpreted as the average change in the probability of a given profile being preferred whenever the given profile displays the attribute level of interest instead of the baseline attribute level (Hainmueller et al., 2014). This way, it is possible to estimate not only the effect of an attribute taking on all its possible values, but also its effect across other possible attributes, which allows the study to explore the possible relative weight that voters may assign to various aspects within their multidimensional choice framework (Hainmueller et al., 2014).

Note that, by design, the choice task outcomes are strongly negatively correlated, because choosing one profile necessarily means not choosing all others, and that the outcomes obtained are mostly driven by unobserved respondent characteristics, who will therefore always choose their preferred combination of attributes whenever they are displayed (Hainmueller et al., 2014). For that reason, when estimating sampling variance, it is important to correct standard errors: this can be done in two ways. First, by calculating cluster-robust standard errors (when population inferences suffer from possible correlated standard errors within, in this case, respondent clusters); or, second, by bootstrapping resampled respondents and then calculating uncertainty estimates with the help of the observed distribution of AMCEs over the resamples (Hainmueller et al., 2014).

# Experimental Design

A pilot study was performed largely influenced by the candidate conjoint analysis performed by Hainmueller et al. (2014), in which hypothetical candidate profiles were shown to respondents for them to choose who they would support, without theorizing about the contextual framework of the election. The study employs a total of nine candidate attributes with a combined number of 45 levels which will be independently and randomly displayed to respondents in a pairwise fashion as a hard-choice task, for a total of 10 choice tasks for respondents (Bansak et al., 2018, 2021). Each respondent will be asked to choose each time between the displayed profiles the one that she would support the most. The pilot test asked 75 respondents for a total *N*= 1500 observations. The survey was created using the survey platform QuestionPro and the survey link was administered to respondents in the United States using the Amazon Mechanical Turk in the form of Human Intelligence Tasks – HITs (Paolacci & Chandler, 2014). The attributes chosen, and their corresponding levels, will be introduced next, along with justification regarding their selection (when considered necessary). The *Ethnicity*, *Occupation*, *Gender*, *Income* and *Age* attributes are based the candidates experiment in Hainmueller et al. (2014).

Party Affiliation

The study introduces a third level 'Party Identification not available' to be displayed with a probability of 0.66, to avoid generalized strong party identification effects in responses, which would most likely arise from the marked partisanship that characterizes American politics (Buttice & Stone, 2012; Gouret et al., 2011; Palmer et al., 2013; Sanders et al., 2011). The levels are:

1. Republican
2. Democrat
3. Party Identification not available

Ethnicity

1. African American
2. Hispanic/Latino
3. White non-Hispanic
4. Native/American
5. Asian

Incumbency status

Another critical valence attribute in American politics (Hainmueller & Kern, 2008; Levitt, 1994; Levitt & Wolfram, 1997; Stone & Simas, 2010), the wording of the two levels has been simplified in order to be less technical and to not contain specialized words such as *incumbent* or *challenger*. The levels are:

1. The candidate is in office and seeks reelection.
2. The candidate is looking to be elected for the first time.

Gender

Following Mcdermott (1997), the inclusion of a gender variable responds to the intention of interacting it for possible gender effects between respondents' and candidates' attributes. The levels are:

1. Female
2. Male

Occupation

Based on the immigration experiment in Hainmueller et al. (2014), but also inspired in the notion of famous political amateurs entering electoral races. Levels such as *actor* or *athlete* play, therefore, an important role in the experiment, especially when potentially interacted with high levels of public prominence. Furthermore, it has been recently found that in local races voters tend to support candidates with a previous occupation related to the office they are running for (Atkeson & Hamel, 2020). The attributes are:

1. Lawyer
2. Military Officer
3. Teacher
4. Farmer
5. Business owner
6. Athlete
7. Actor
8. Banker
9. Journalist
10. Union Leader

Age

Recent evidence indicates that age is indeed an important and understudied heuristic through which voters tend to favor candidates with the same party affiliation that are closer in age to them (Webster & Pierce, 2019). The levels are:

1. 31 years old
2. 38 years old
3. 45 years old
4. 52 years old
5. 59 years old
6. 66 years old

7. 73 years old

Income

A reference for this attribute was Wüest & Rosset (2017) in the Swiss case. The levels consist of the following fixed annual income figures that represent incomes rising to just short of the top 1% (Winters & Page, 2009):

1. Annual income $32,000
2. Annual income $54,000
3. Annual income $75,000
4. Annual income $92,000
5. Annual income $140,000
6. Annual income $360,000
7. Annual income $840,000

Public Prominence

Based on the notion of prominence as a component of the importance of political actors (Munzert, 2018). Public prominence will be operationalized and signaled to respondents using a 'followers on Twitter' metric. This metric is a good operationalization of the idea of prominence, since it gives the respondent an idea of how many others are dedicating time of their own to follow the candidate on social media, meaning that, naturally, the more followers a candidate counts with, the more prominent she is. The levels were designed as follows:

1. The candidate has 210 followers on Twitter

2. The candidate has 2.400 followers on Twitter

3. The candidate has 23.700 followers on Twitter

4. The candidate has 315.000 followers on Twitter

5. The candidate has 1.3 million followers on Twitter

Political Prominence

Operationalized and signaled to respondents indicating whether the candidate played an important role in her political party. The levels are the following:

1. The candidate has not played a major role in the party
2. The candidate is a locally renowned member of the party
3. The candidate is a state-wide renowned member of the party
4. The candidate is a nationally renowned member of the party

Respondents survey questions

After the 10 choice tasks presented to the respondent, they were asked to answer seven mandatory sociodemographic and political ideological self-placement questions. The questions are the following:

a) In a scale from 0 to 10, where 0 indicates "Far Left" and 10 indicates "Far Right", where would you place yourself in terms of political ideology support?

b) Select from of the following ethnicities, the one with which you identify yourself most:

   i. White

   ii. African American

   iii. Hispanic/ Latinx

   iv. Asian

   v. Native American

   vi. Prefer not to say

c) Select from the following age ranges, the one in which you are located:

   i. Younger than 20 years old

   ii. 20 - 30 years old

   iii. 31 - 40 years old

   iv. 41 - 50 years old

   v. 51 - 60 years old

   vi. 61- 70 years old

   vii. Older than 70 years old

d) Do you usually think of yourself as a Republican, a Democrat, an Independent, or something else?

   i. Republican

   ii. Democrat

   iii. Independent

   iv. Something else

e) In general terms: How interested in politics are you?

   i. Not interested at all

ii. Slightly interested

iii. Moderately interested

iv. Rather interested

v. Very interested

d) What is your gender?

i. Male

ii. Female

iii. Other

iv. Prefer not to say

e) What is the highest level of school you have completed?

i. No schooling

ii. Some high school, no diploma

iii. High school diploma or equivalent

iv. Some college or university studies, not completed

v. College or university studies, completed

vi. Graduate studies

# Results of Preliminary Analysis

After running the AMCE model with the help of the *cjoint* package in the statistical software R, we found a significant positive effect of the perceptions of political prominence on the probability of the candidate being elected, meaning that for RQ1, levels of perceived political prominence do matter by themselves. Figure 2 presents all the AMCEs as component change in the expected probability of a profile being chosen for every attribute and every level, compared to a baseline level, all else held equal. See Table 1 in the Appendix for detailed AMCE estimates.

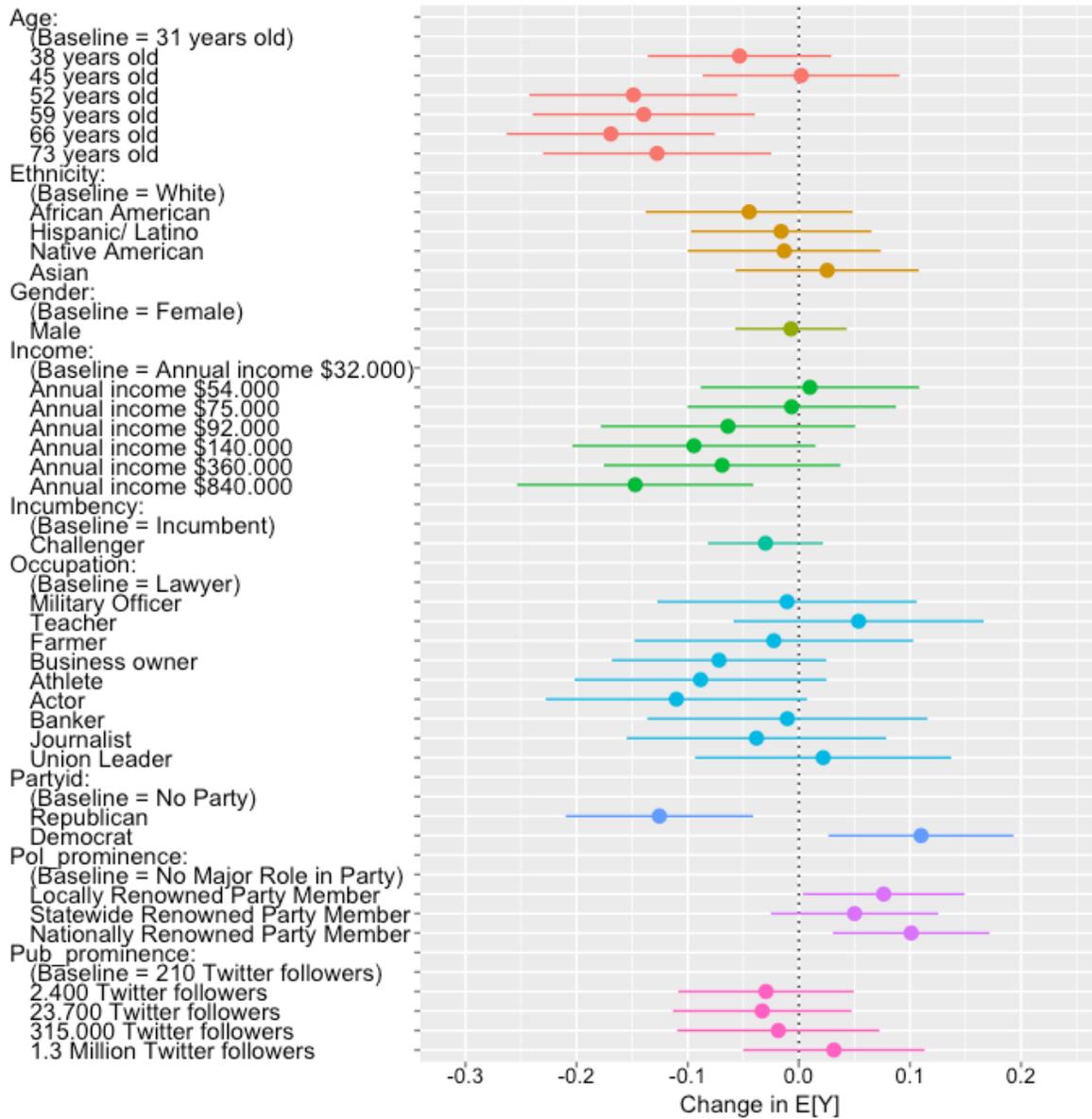

**Figure 2** – AMCE estimates for all attributes and levels

The marked effect of partisanship of the candidate is likely due to the high imbalance in the respondent sample, with Democrat respondents accounting for close to 55%, Republicans for the 17% and Independents for 27% of the total. The latter explains why belonging to the Republican party or having the highest level of income on average decreases the probability of the candidate being elected. However, the more interesting result pertaining to the study refers to the significant and positive effect of political prominence on the probability of election (except for the non-significant estimate for "Statewide Renowned Party Member"). We theorize that the recognized levels of prominence depends greatly on the subjective informational structure of the voter and therefore the significant positive effects of political prominence might be due to the response distribution regarding how interested they said they are in politics, which is slightly skewed in the direction of overall greater interest in politics: Not interested at all, 5%; slightly interested, 20%; moderately interested, 32%; rather interested, 27%; very interested, 16%.

To further explore why this might be the case and to answer RQ2 and RQ3, we propose a conditional estimation of prominence regarding respondents' characteristics. Figure 3 shows ACIE estimations of perceptions of political and public prominence, along with the candidate's political party affiliation, on a grid that varies along the 11-point scale of the respondent's own ideological self-placement, ranging from 0 – 'Far left' to 10 – 'Far right'. As expected, it is possible to see how the effect of the candidate's party shifts as the scale increases, from a more positive support for Democratic candidates at the beginning towards more positive support for the Republican candidate, as compared to a candidate with no party information displayed. Drawing attention only to the change in the effect of perceptions of political prominence of candidates, conditional on the ideological self-placement of the respondents, there is not too much of a gradual shift. The same could be said about the effect of perceptions of public prominence conditional on the ideological stance of the respondents. Although there are specific cases in which the effect may be clearly positive or negative and, above all, statistically significant, it is not possible to say that these changes in estimates are a function of the ideological position of the respondent. In general, estimates are non-significant, with some exceptions for instance when respondents locate themselves in position 2 and 6 of the ideological scale. In those cases, political prominence seems to keep having the general positive effect on the probability of candidate selection. However, for respondents who locate themselves at the rightmost position of the scale (lower right corner plot in Figure 3, position 11 on the scale), political prominence acquires a more negative effect on the probability of selection, whereas public prominence acquires a positive effect. Table 2 in the Appendix details the ACIEs for this last interaction estimation.

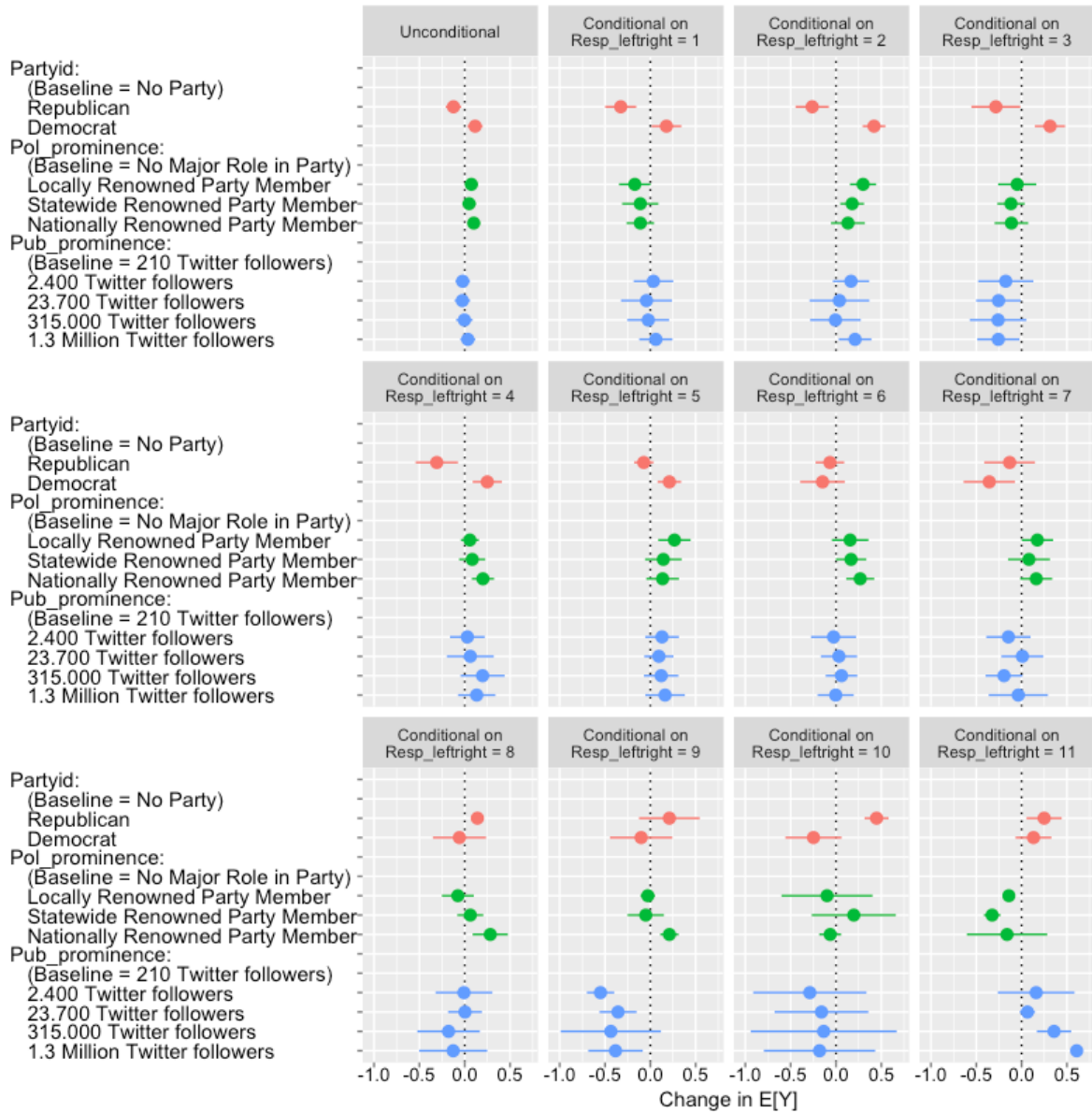

**Figure 3** - Conditional effects of a candidate's PartyID and Public and Political Prominence on respondents' left-right ideological self-placement

Figure 4 shows the effects of the candidates' attributes, conditional on the respondents' self-assessed interest in politics. See Tables 3.1 – 3.5 in the Appendix for the complete estimates displayed in Figure 4. In general, respondents who said not to be at all interested in politics showed negative and significant effects for levels of political prominence, particularly for candidates who were locally and nationally renowned members of the party. The latter makes sense, as potential voters who are not familiar with political dynamics and attributable competence or viability associated with different levels of political prominence would not be familiar with how to allocate relative importance to any of the possible signaled scenarios. Moreover, they also showed positive and statistically significant effects for all levels of public prominence, as compared to the baseline (almost no followers on Twitter). This is a major finding, since it supports the idea of publicly perceived prominent political figures being more appealing to those voters who do not care that much about politics. These results would indicate that this group of voters uses a recognition heuristic to make ecologically rational inferences about the viability of the candidate based on the signaled mediator of Twitter followers. Although the proportion of respondents who said not to be interested in politics at all only accounted for 5% of the responses, this group favored further candidates whose occupation was banker and who had higher annual incomes, i.e. $140.000 USD, $360.000 USD and $840.000 USD.

Furthermore, respondents who said to be slightly interested in politics only showed positive statistically significant effects for the candidates who belonged to the Democratic Party. Moderately interested respondents in politics showed positive and statistically significant effects for candidates who tended to lower levels of political prominence, favoring those who were locally or statewide renowned members of the party. They did not show any significant effect towards the candidates' levels of public prominence. Respondents who said they were rather interested in politics showed significant effects regarding the candidates' higher levels of political and public prominence. Specifically, they tended to approve of candidates who were nationally renowned members of the party and tended to disapprove of candidates who had 315.000 Twitter followers. They also showed a negative significant effect for candidates whose annual income was very high ($840.000). Lastly, respondents that said they were very interested in politics did not seem to pay much attention to the cues and signals relating to levels of perceived political and

public prominence. This finding would support the idea of voters being less prone to use heuristics when they are (very) interested in politics, as opposed to those who are not, as mentioned above.

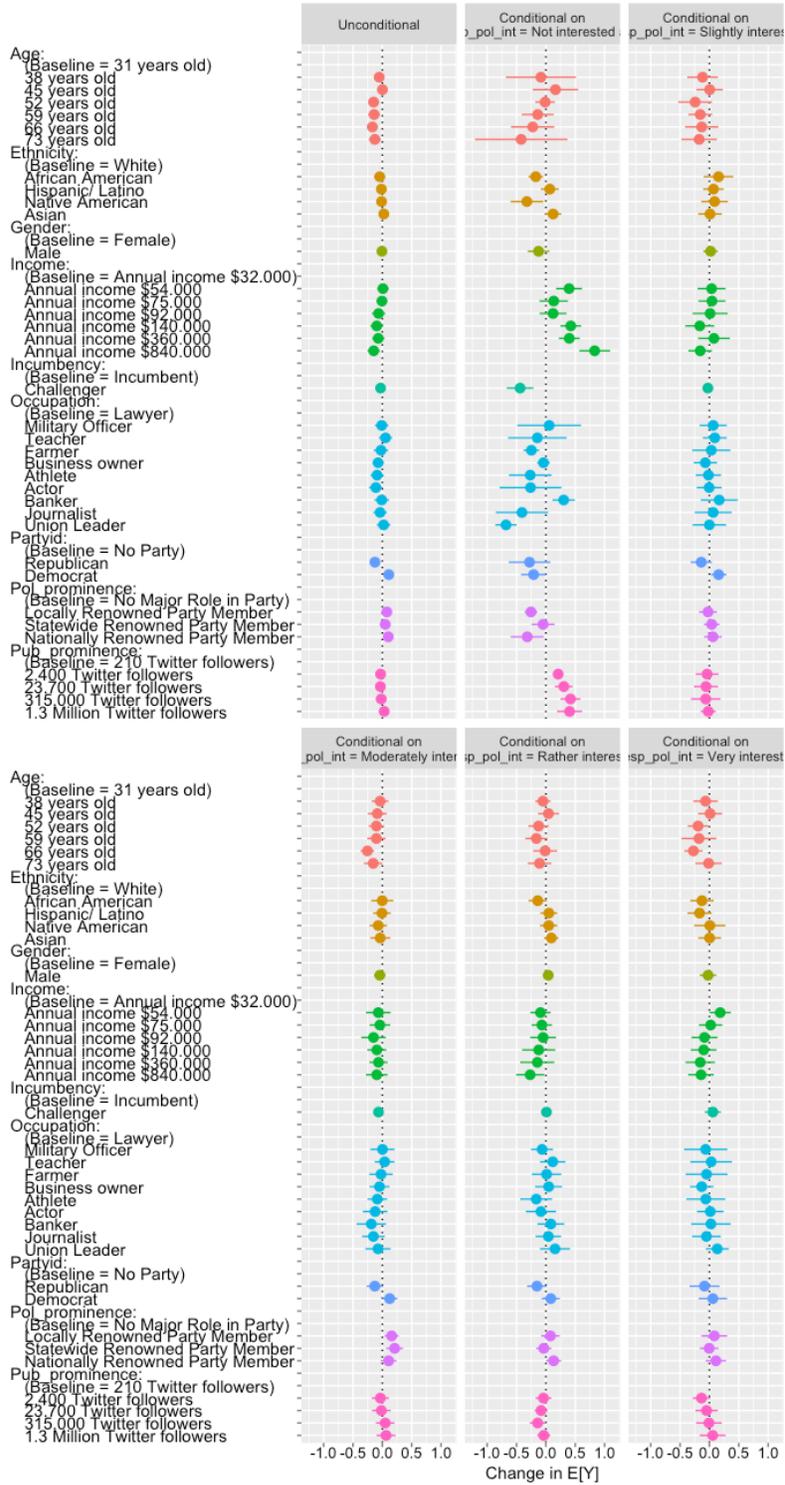

**Figure 4** - Conditional effects of a candidate's attributes on respondents' self-assessed interest in politics

Finally, and to explore how education levels might interact with the effect of prominence on voting, Figure 5 shows the conditional effects of a candidate's attributes on different levels of respondents' education. See Tables 4.1 – 4.5 in the Appendix for the complete estimates displayed in Figure 5.

Although representing only the 3% of the responses, respondents who did some high school but never finished show highly significant effects for all candidates' levels of public and political prominence. All those effects of the perceived prominence of candidates are positive for this group of prospective voters, except for a candidate who is a statewide renowned member of the party, which has a negative effect. Those respondents who did some college or university studies but never completed them (21% of the responses) showed a statistically significant positive effect for higher levels of political prominence of candidates, particularly for candidates who were nationally renowned members of the party. Respondents who finished their undergraduate education (56% of the responses) showed positive significant effects for candidates that were 45 years old and for candidates who belonged to the Democratic Party. Finally, those respondents who did some graduate studies (5% of the responses) showed a statistically significant preference towards candidates with higher levels of political prominence and lower levels of public prominence. Specifically, they showed positive effects for candidates who were nationally renowned members of the party and for those who had 23.700 Twitter Followers. This finding would suggest that highly educated respondents rely on high political prominence levels and on lower levels of public prominence as recognition heuristics at the time of voting. This group of prospective voters, tended to dislike Republican candidates.

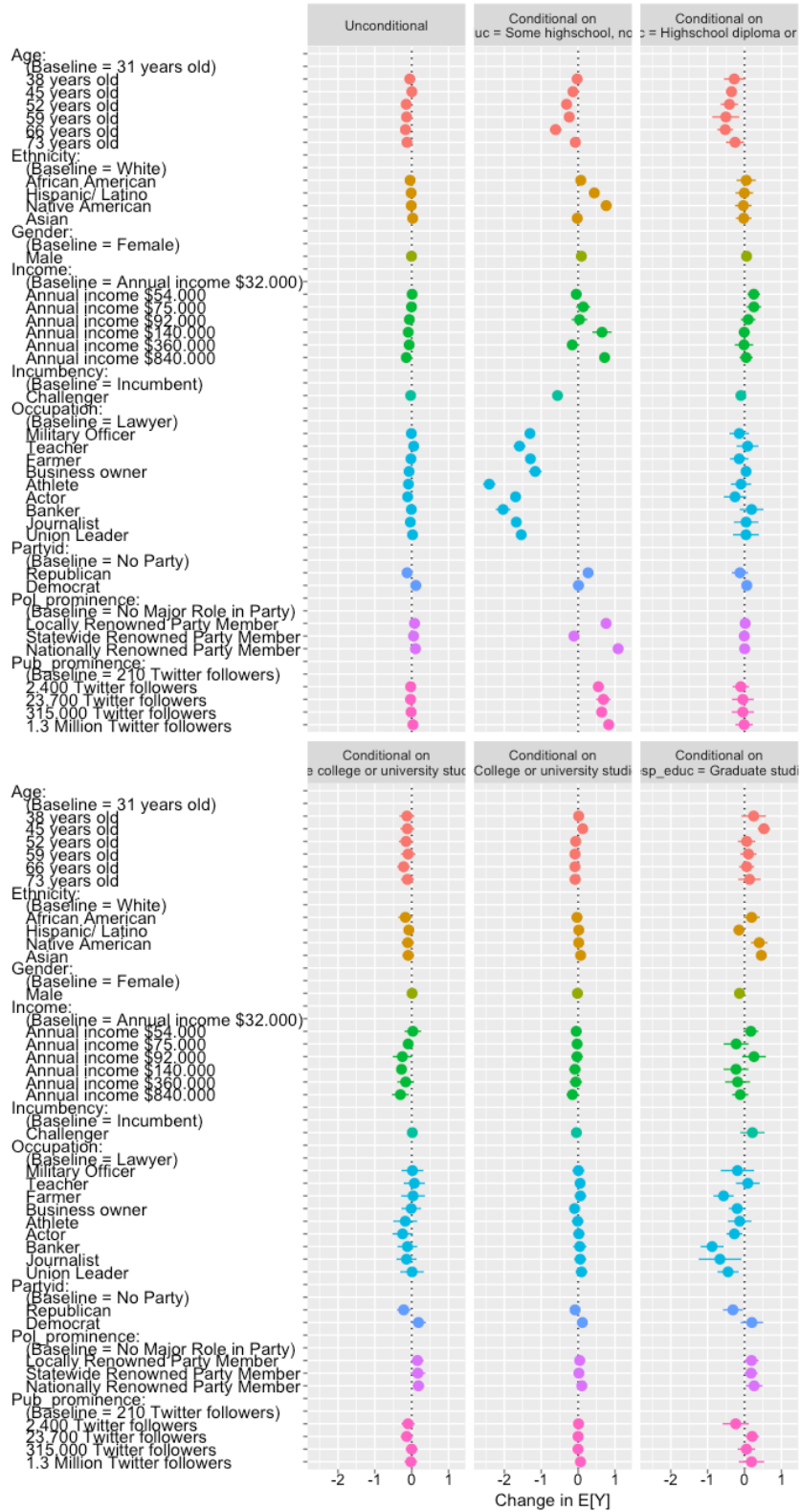

**Figure 5 -** Conditional effects of a candidate's attributes on respondents' level of education

The results indicate significant effects of different levels of political or public prominence, by themselves or conditioned on respondent varying characteristics. Nevertheless, the issue remains one of replicability and validity, especially as the pilot tests presented might be underpowered. By design, the experimental conjoint analysis allows for significant internal validity of the model (Hainmueller et al., 2014). Regarding external validity, however, further attempts via replication of the experiment with a larger sample and comparative research is warranted (Hainmueller et al., 2015). Theoretical explanations to the patterns and effects found can be manifold, and thus must be treated with care to avoid premature causal inferences, although initial findings regarding all three research questions seem to be satisfying. Finally, a caveat of our design must be put forward: due to technical limitations of the online survey platform used, the randomly displayed attribute levels were shown in a fixed order every time. This aspect could have influenced the estimates by inducing primacy effects on the respondents (Hainmueller et al., 2014).

These findings directly call for the crucial need to understand the heuristics that drive electoral races and help shape their outcomes. The existing literature on valence politics has shed much light on the matter, and now this research adds another piece to the puzzle, further highlighting the complex dynamics at play in electoral races. The importance of understanding said heuristics is however much needed to further design and update electoral systems with the aim of avoiding possible system abuses or even electoral related manipulations, like the possibility of spending large amounts of money on advertising. This with the goal of making a candidate more publicly prominent and making her more appealing to the share of voting population that is not interest in politics, but even more importantly, for those who do not vote (estimated at over 40% of the voting-age population (Desilver, 2020)) and for which a lack of interest in politics may be a rather important incentive not to do so. Further avenues of research and possible policy implications could relate to ballot redesign with the aim of getting ahead of possibly dangerous heuristics voters use and to disentangle why right-wing extremists apparently place more importance on public perceptions of prominence and a negative weight on political prominence.

# References

Ansolabehere, S., Snyder, J. M., & Stewart, C. (2001). Candidate Positioning in U.S. House Elections. *American Journal of Political Science*, *45*(1), 136. https://doi.org/10.2307/2669364

Atkeson, L. R., & Hamel, B. T. (2020). Fit for the Job: Candidate Qualifications and Vote Choice in Low Information Elections. *Political Behavior*, *42*(1), 59–82. https://doi.org/10.1007/S11109-018-9486-0/FIGURES/4

Bansak, K., Hainmueller, J., Hopkins, D. J., & Yamamoto, T. (2018). The Number of Choice Tasks and Survey Satisficing in Conjoint Experiments. *Political Analysis*, *26*(1), 112–119. https://doi.org/10.1017/pan.2017.40

Bansak, K., Hainmueller, J., Hopkins, D. J., & Yamamoto, T. (2021). Beyond the breaking point? Survey satisficing in conjoint experiments. *Political Science Research and Methods*, *9*(1), 53–71. https://doi.org/10.1017/psrm.2019.13

Boas, D. T. C. (2014). Pastor Paulo vs. Doctor Carlos: Professional Titles as Voting Heuristics in Brazil: *Https://Doi.Org/10.1177/1866802X1400600202*, *6*(2), 39–72. https://doi.org/10.1177/1866802X1400600202

Buttice, M. K., & Stone, W. J. (2012). Candidates matter: Policy and quality differences in congressional elections. *Journal of Politics*, *74*(3), 870–887. https://doi.org/10.1017/S0022381612000394

Carnes, N., & Lupu, N. (2016). Do voters dislike working-class candidates? Voter biases and the descriptive underrepresentation of the working class. *American Political Science Review*, *110*(4),

832–844. https://doi.org/10.1017/S0003055416000551

Desilver, D. (2020). In past elections, U.S. trailed most developed countries in voter turnout _ Pew Research Center. *Pew Research Center*. https://www.pewresearch.org/fact-tank/2020/11/03/in-past-elections-u-s-trailed-most-developed-countries-in-voter-turnout/

Downs, A. (1957). *An Economic Theory of Democracy*. Harper and Brothers.

Franchino, F., & Zucchini, F. (2015). Voting in a Multi-dimensional Space: A Conjoint Analysis Employing Valence and Ideology Attributes of Candidates. *Political Science Research and Methods*, *3*(2), 221–241. https://doi.org/10.1017/psrm.2014.24

Gaissmaier, W., & Marewski, J. N. (2011). Forecasting elections with mere recognition from small, lousy samples: A comparison of collective recognition, wisdom of crowds, and representative polls. *Judgment and Decision Making*, *6*(1), 73–88. http://nbn-resolving.de/urn:nbn:de:bsz:352-279320

Goldstein, D. G., & Gigerenzer, G. (2002). Models of ecological rationality: The recognition heuristic. *Psychological Review*, *109*(1), 75–90. https://doi.org/10.1037/0033-295x.109.1.75

Gouret, F., Hollard, G., & Rossignol, S. (2011). An empirical analysis of valence in electoral competition. *Social Choice and Welfare*, *37*(2), 309–340. https://doi.org/10.1007/s00355-010-0495-0

Hainmueller, J., Hangartner, D., & Yamamoto, T. (2015). Validating vignette and conjoint survey experiments against real-world behavior. *Proceedings of the National Academy of Sciences of the United States of America*, *112*(8), 2395–2400. https://doi.org/10.1073/pnas.1416587112

Hainmueller, J., & Hopkins, D. J. (2014). Public attitudes toward immigration. *Annual Review of Political Science*, *17*, 225–249. https://doi.org/10.1146/annurev-polisci-102512-194818

Hainmueller, J., Hopkins, D. J., & Yamamoto, T. (2014). Causal inference in conjoint analysis: Understanding multidimensional choices via stated preference experiments. *Political Analysis*, *22*(1), 1–30. https://doi.org/10.1093/pan/mpt024

Hainmueller, J., & Kern, H. L. (2008). Incumbency as a source of spillover effects in mixed electoral systems: Evidence from a regression-discontinuity design. *Electoral Studies*, *27*(2), 213–227. https://doi.org/10.1016/j.electstud.2007.10.006

Kam, C. D., & Zechmeister, E. J. (2013). Name Recognition and Candidate Support. *American Journal of Political Science*, *57*(4), 971–986. https://doi.org/10.1111/ajps.12034

Levitt, S. D. (1994). Using Repeat Challengers to Estimate the Effect of Campaign Spending on Election Outcomes in the U.S. House. *Journal of Political Economy*, *102*(4), 777–798. https://doi.org/10.1086/261954

Levitt, S. D., & Wolfram, C. D. (1997). Decomposing the Sources of Incumbency Advantage in the U. S. House. *Legislative Studies Quarterly*, *22*(1), 45. https://doi.org/10.2307/440290

Mcdermott, M. L. (1997). Voting Cues in Low-Information Elections: Candidate Gender as a Social Information Variable in Contemporary United States Elections. *Source: American Journal of Political Science*, *41*(1), 270–283.

McGregor, S. C. (2017). Personalization, social media, and voting: Effects of candidate self-personalization on vote intention: *Https://Doi.Org/10.1177/1461444816686103*, *20*(3), 1139–1160. https://doi.org/10.1177/1461444816686103

Munzert, S. (2018). *Measuring the Importance of Political Elites*. https://doi.org/10.31235/osf.io/t8gs5

Palmer, H. D., Whitten, G. D., & Williams, L. K. (2013). Who should be chef? The dynamics of valence evaluations across income groups during economic crises. *Electoral Studies*, *32*(3), 425–

431. https://doi.org/10.1016/j.electstud.2013.05.014

Panagopoulos, C., & Green, D. P. (2008). Field experiments testing the impact of radio advertisements on electoral competition. *American Journal of Political Science*, *52*(1), 156–168. https://doi.org/10.1111/j.1540-5907.2007.00305.x

Paolacci, G., & Chandler, J. (2014). Inside the Turk: Understanding Mechanical Turk as a Participant Pool. *Current Directions in Psychological Science*, *23*(3), 184–188. https://doi.org/10.1177/0963721414531598

Sanders, D., Clarke, H. D., Stewart, M. C., & Whiteley, P. (2011). Downs, Stokes and the Dynamics of Electoral Choice. *British Journal of Political Science*, *41*(2), 287–314. https://doi.org/10.1017/S0007123410000505

Simon, H. A. (1955). A Behavioral Model of Rational Choice. *Source: The Quarterly Journal of Economics*, *69*(1), 99–118. https://about.jstor.org/terms

Stokes, D. E. (1963). Spatial Models of Party Competition. *American Political Science Review*, *57*(2), 368–377. https://doi.org/10.2307/1952828

Stone, W. J., & Simas, E. N. (2010). Candidate Valence and Ideological Positions in U.S. House Elections. *American Journal of Political Science*, *54*(2), 371–388. https://doi.org/10.1111/j.1540-5907.2010.00436.x

Webster, S. W., & Pierce, A. W. (2019). Older, Younger, or More Similar? The Use of Age as a Voting Heuristic*. *Social Science Quarterly*, *100*(3), 635–652. https://doi.org/10.1111/SSQU.12604

Winters, J. A., & Page, B. I. (2009). Oligarchy in the United States? *Perspectives on Politics*, *7*(4), 731–751. https://doi.org/10.1017/S1537592709991770

Wüest, R., & Rosset, J. (2017). Legislator Income and Voting Behavior. *SSRN Electronic Journal*. https://doi.org/10.2139/ssrn.3077584

# Appendix

Table 1 – *General AMCEs*

```
------------------------------------------
Average Marginal Component Effects (AMCE):
------------------------------------------
    Attribute                           Level   Estimate  Std. Err    z value    Pr(>|z|)
          Age                    38 years old -0.0533221  0.042160  -1.264765  0.20595575
          Age                    45 years old  0.0020436  0.045214   0.045198  0.96394917
          Age                    52 years old -0.1488586  0.047792  -3.114732  0.00184112 **
          Age                    59 years old -0.1395189  0.050980  -2.736720  0.00620550 **
          Age                    66 years old -0.1692445  0.047813  -3.539742  0.00040052 ***
          Age                    73 years old -0.1275655  0.052357  -2.436450  0.01483222 *
    Ethnicity               African American -0.0446866  0.047551  -0.939759  0.34734101
    Ethnicity                Hispanic/ Latino -0.0159455  0.041431  -0.384873  0.70033159
    Ethnicity                 Native American -0.0131381  0.044343  -0.296281  0.76701582
    Ethnicity                           Asian  0.0255200  0.042089   0.606341  0.54428830
       Gender                            Male -0.0070693  0.025608  -0.276058  0.78250316
       Income            Annual income $54.000  0.0100045  0.050200   0.199293  0.84203353
       Income            Annual income $75.000 -0.0064426  0.047856  -0.134624  0.89290932
       Income            Annual income $92.000 -0.0637552  0.058369  -1.092284  0.27470844
       Income           Annual income $140.000 -0.0943947  0.055877  -1.689316  0.09115886
       Income           Annual income $360.000 -0.0690195  0.054306  -1.270933  0.20375255
       Income           Annual income $840.000 -0.1472766  0.054215  -2.716534  0.00659694 **
   Incumbency                        Challenger -0.0300054  0.026401  -1.136523  0.25573769
   Occupation                  Military Officer -0.0106141  0.059565  -0.178193  0.85857133
   Occupation                           Teacher  0.0538062  0.057436   0.936804  0.34885954
   Occupation                            Farmer -0.0225060  0.064078  -0.351230  0.72541569
   Occupation                    Business owner -0.0717899  0.049232  -1.458185  0.14478957
   Occupation                           Athlete -0.0884233  0.057840  -1.528744  0.12632786
   Occupation                             Actor -0.1101798  0.060019  -1.835756  0.06639379
   Occupation                             Banker -0.0103466  0.064381  -0.160707  0.87232381
   Occupation                         Journalist -0.0381054  0.059569  -0.639685  0.52237752
   Occupation                        Union Leader  0.0218979  0.058825   0.372257  0.70970153
      Partyid                          Republican -0.1255247  0.043001  -2.919132  0.00351007 **
      Partyid                            Democrat  0.1100048  0.042502   2.588198  0.00964796 **
Pol_prominence    Locally Renowned Party Member  0.0763782  0.037079   2.059880  0.03941003 *
Pol_prominence   Statewide Renowned Party Member  0.0501424  0.038406   1.305574  0.19169753
Pol_prominence Nationally Renowned Party Member  0.1011408  0.035927   2.815200  0.00487469 **
Pub_prominence            2.400 Twitter followers -0.0295732  0.040258  -0.734592  0.46258780
Pub_prominence           23.700 Twitter followers -0.0329326  0.041040  -0.802460  0.42228694
Pub_prominence          315.000 Twitter followers -0.0184331  0.046373  -0.397493  0.69100383
Pub_prominence       1.3 Million Twitter followers  0.0315766  0.041677   0.757647  0.44866234
---
Number of Obs. = 1500
---
Number of Respondents = 75
---
Signif. codes: 0 '***' 0.001 '**' 0.01 '*' 0.05

--------------------
```

Table 2 – *Conditional effects of a candidate's PartyID and Public and Political Prominence on respondents' left-right ideological self-placement on the far right.*

```
------------------------------------------------------------
Conditional AMCE's (Respleftright = 11):
------------------------------------------------------------
       Attribute                           Level  Estimate Std. Err   z value       Pr(>|z|)
 Pol_prominence     Locally Renowned Party Member -0.140778 0.027664 -5.08879  3.6035e-07 ***
 Pol_prominence   Statewide Renowned Party Member -0.325025 0.047457 -6.84884  7.4453e-12 ***
 Pol_prominence  Nationally Renowned Party Member -0.160947 0.226247 -0.71138  4.7685e-01
 Pub_prominence             2.400 Twitter followers  0.160376 0.216134  0.74202  4.5807e-01
 Pub_prominence            23.700 Twitter followers  0.063971 0.043227  1.47989  1.3890e-01
 Pub_prominence           315.000 Twitter followers  0.357559 0.096765  3.69512  2.1978e-04 ***
 Pub_prominence       1.3 Million Twitter followers  0.607424 0.020612 29.46936 7.1120e-191 ***
         Partyid                        Republican  0.248557 0.098913  2.51288  1.1975e-02   *
         Partyid                          Democrat  0.131088 0.100952  1.29852  1.9411e-01
---
Number of Obs. = 1500
Number of Respondents = 75
---
Signif. codes: 0 '***' 0.001 '**' 0.01 '*' 0.05

Signif. codes: 0 '***' 0.001 '**' 0.01 '*' 0.05
```

```
-------------------------------------------------------------
Conditional AMCE's (Resppartisanship = Something else):
-------------------------------------------------------------
       Attribute                        Level Estimate Std. Err   z value    Pr(>|z|)
   Pub_prominence         2.400 Twitter followers  0.37736  0.18992   1.98695 4.6928e-02   *
   Pub_prominence        23.700 Twitter followers  3.67201  0.77894   4.71411 2.4276e-06 ***
   Pub_prominence       315.000 Twitter followers  0.78992  0.26648   2.96432 3.0335e-03  **
   Pub_prominence   1.3 Million Twitter followers  0.87699  0.40972   2.14045 3.2318e-02   *
   Pol_prominence    Locally Renowned Party Member -0.81007  0.20251  -4.00017 6.3298e-05 ***
   Pol_prominence  Statewide Renowned Party Member -0.56244  0.29281  -1.92087 5.4748e-02
   Pol_prominence Nationally Renowned Party Member -1.32071  0.35540  -3.71614 2.0229e-04 ***
           Income          Annual income $54.000   1.17348  0.59447   1.97397 4.8385e-02   *
           Income          Annual income $75.000  -1.40057  1.21009  -1.15741 2.4711e-01
           Income          Annual income $92.000  -1.69345  1.03079  -1.64286 1.0041e-01
           Income         Annual income $140.000  -0.71805  0.90489  -0.79352 4.2747e-01
           Income         Annual income $360.000  -1.65933  1.14460  -1.44971 1.4714e-01
           Income         Annual income $840.000  -0.64494  0.97980  -0.65823 5.1039e-01
              Age                  38 years old  -0.34836  0.53224  -0.65452 5.1278e-01
              Age                  45 years old   2.50085  0.49818   5.01999 5.1674e-07 ***
              Age                  52 years old   0.61524  0.66268   0.92841 3.5320e-01
              Age                  59 years old  -2.08102  0.30357  -6.85507 7.1279e-12 ***
              Age                  66 years old        NA       NA        NA         NA
              Age                  73 years old   1.81624  0.37298   4.86959 1.1183e-06 ***
       Occupation              Military Officer   1.56169  0.44458   3.51278 4.4345e-04 ***
       Occupation                       Teacher        NA       NA        NA         NA
       Occupation                        Farmer        NA       NA        NA         NA
       Occupation                Business owner        NA       NA        NA         NA
       Occupation                       Athlete        NA       NA        NA         NA
       Occupation                         Actor        NA       NA        NA         NA
       Occupation                        Banker        NA       NA        NA         NA
       Occupation                    Journalist        NA       NA        NA         NA
       Occupation                  Union Leader        NA       NA        NA         NA
           Gender                          Male        NA       NA        NA         NA
       Incumbency                    Challenger        NA       NA        NA         NA
        Ethnicity              African American        NA       NA        NA         NA
        Ethnicity               Hispanic/ Latino        NA       NA        NA         NA
        Ethnicity               Native American        NA       NA        NA         NA
        Ethnicity                         Asian        NA       NA        NA         NA
          Partyid                    Republican        NA       NA        NA         NA
          Partyid                      Democrat        NA       NA        NA         NA
---
Number of Obs. = 1500
Number of Respondents = 75
---
Signif. codes: 0 '***' 0.001 '**' 0.01 '*' 0.05
```

Table 3.1 - *Conditional effects of a candidate's attributes on respondents' self-assessed interest in politics (Not interested at all)*

```
------------------------------------------------------------
Conditional AMCE's (Resppolint = Not interested at all):
------------------------------------------------------------
        Attribute                          Level  Estimate Std. Err   z value    Pr(>|z|)
   Pol_prominence    Locally Renowned Party Member -0.253597 0.055192 -4.59481 4.3314e-06 ***
   Pol_prominence  Statewide Renowned Party Member -0.048030 0.100309 -0.47882 6.3207e-01
   Pol_prominence Nationally Renowned Party Member -0.317823 0.143461 -2.21540 2.6733e-02   *
   Pub_prominence              2.400 Twitter followers  0.209264 0.041220  5.07673 3.8399e-07 ***
   Pub_prominence             23.700 Twitter followers  0.305581 0.077386  3.94878 7.8552e-05 ***
   Pub_prominence            315.000 Twitter followers  0.419367 0.086711  4.83638 1.3223e-06 ***
   Pub_prominence        1.3 Million Twitter followers  0.404105 0.108514  3.72400 1.9609e-04 ***
           Income             Annual income $54.000  0.395343 0.112940  3.50047 4.6444e-04 ***
           Income             Annual income $75.000  0.134074 0.124404  1.07773 2.8115e-01
           Income             Annual income $92.000  0.122591 0.117022  1.04759 2.9483e-01
           Income            Annual income $140.000  0.424072 0.089714  4.72692 2.2795e-06 ***
           Income            Annual income $360.000  0.399009 0.091505  4.36052 1.2975e-05 ***
           Income            Annual income $840.000  0.831041 0.133733  6.21418 5.1592e-10 ***
              Age                     38 years old -0.084494 0.304702 -0.27730 7.8155e-01
              Age                     45 years old  0.163276 0.197417  0.82706 4.0820e-01
              Age                     52 years old -0.012475 0.085079 -0.14663 8.8343e-01
              Age                     59 years old -0.139460 0.139443 -1.00012 3.1725e-01
              Age                     66 years old -0.224429 0.188046 -1.19348 2.3268e-01
              Age                     73 years old -0.420834 0.402098 -1.04660 2.9529e-01
       Occupation                  Military Officer  0.055882 0.276962  0.20177 8.4010e-01
       Occupation                           Teacher -0.145770 0.254424 -0.57294 5.6669e-01
       Occupation                            Farmer -0.246308 0.069613 -3.53824 4.0281e-04 ***
       Occupation                    Business owner -0.040324 0.055584 -0.72545 4.6818e-01
       Occupation                           Athlete -0.266293 0.185690 -1.43407 1.5155e-01
       Occupation                             Actor -0.261386 0.269105 -0.97132 3.3139e-01
       Occupation                             Banker  0.302483 0.097228  3.11106 1.8642e-03  **
       Occupation                        Journalist -0.407490 0.227681 -1.78974 7.3496e-02
       Occupation                       Union Leader -0.679430 0.092606 -7.33680 2.1876e-13 ***
           Gender                              Male -0.124127 0.094878 -1.30828 1.9078e-01
       Incumbency                         Challenger -0.438860 0.116231 -3.77577 1.5952e-04 ***
        Ethnicity                   African American -0.171942 0.063213 -2.72003 6.5277e-03  **
        Ethnicity                    Hispanic/ Latino  0.066530 0.077542  0.85798 3.9090e-01
        Ethnicity                    Native American -0.324434 0.141814 -2.28774 2.2153e-02   *
        Ethnicity                              Asian  0.122972 0.071672  1.71576 8.6206e-02
          Partyid                         Republican -0.278275 0.180845 -1.53875 1.2387e-01
          Partyid                           Democrat -0.209559 0.109105 -1.92071 5.4768e-02
---
Number of Obs. = 1500
Number of Respondents = 75
---
Signif. codes: 0 '***' 0.001 '**' 0.01 '*' 0.05
```

Table 3.2 - *Conditional effects of a candidate's attributes on respondents' self-assessed interest in politics (Slightly interested)*

```
-------------------------------------------------------------
Conditional AMCE's (Resppolint = Slightly interested):
-------------------------------------------------------------
      Attribute                                Level   Estimate Std. Err    z value Pr(>|z|)
  Pol_prominence     Locally Renowned Party Member  -0.0236361 0.078719 -0.3002596 0.763979
  Pol_prominence    Statewide Renowned Party Member   0.0385352 0.068005  0.5666540 0.570949
  Pol_prominence   Nationally Renowned Party Member   0.0591382 0.077573  0.7623591 0.445846
  Pub_prominence              2.400 Twitter followers -0.0387382 0.099073 -0.3910079 0.695791
  Pub_prominence             23.700 Twitter followers -0.0579076 0.104858 -0.5522503 0.580777
  Pub_prominence            315.000 Twitter followers -0.0625174 0.128112 -0.4879906 0.625556
  Pub_prominence        1.3 Million Twitter followers -0.0204148 0.065892 -0.3098234 0.756695
         Income              Annual income $54.000   0.0372293 0.121995  0.3051709 0.760236
         Income              Annual income $75.000   0.0430308 0.118647  0.3626781 0.716845
         Income              Annual income $92.000   0.0115412 0.152864  0.0754999 0.939817
         Income             Annual income $140.000  -0.1649070 0.126423 -1.3044095 0.192094
         Income             Annual income $360.000   0.0764424 0.138215  0.5530681 0.580217
         Income             Annual income $840.000  -0.1570707 0.102543 -1.5317554 0.125583
            Age                       38 years old  -0.1174582 0.133227 -0.8816413 0.377971
            Age                       45 years old   0.0042222 0.114107  0.0370025 0.970483
            Age                       52 years old  -0.2435596 0.147211 -1.6544931 0.098027
            Age                       59 years old  -0.1564726 0.103395 -1.5133450 0.130192
            Age                       66 years old  -0.1314714 0.144218 -0.9116150 0.361971
            Age                       73 years old  -0.1741361 0.153557 -1.1340141 0.256789
     Occupation                   Military Officer   0.0640380 0.118426  0.5407412 0.588686
     Occupation                            Teacher   0.0910795 0.104315  0.8731226 0.382596
     Occupation                             Farmer   0.0318313 0.166527  0.1911482 0.848410
     Occupation                     Business owner  -0.0687890 0.101735 -0.6761598 0.498939
     Occupation                            Athlete  -0.0161696 0.110561 -0.1462506 0.883724
     Occupation                              Actor  -0.0043981 0.108610 -0.0404939 0.967699
     Occupation                             Banker   0.1664791 0.160925  1.0345114 0.300897
     Occupation                         Journalist   0.0641756 0.160979  0.3986570 0.690146
     Occupation                       Union Leader  -0.0013217 0.146311 -0.0090335 0.992792
         Gender                               Male   0.0194894 0.063835  0.3053067 0.760133
     Incumbency                         Challenger  -0.0266102 0.044619 -0.5963868 0.550917
      Ethnicity                   African American   0.1549202 0.129163  1.1994172 0.230366
      Ethnicity                    Hispanic/ Latino  0.0685817 0.089776  0.7639174 0.444916
      Ethnicity                    Native American   0.0870159 0.115606  0.7526964 0.451632
      Ethnicity                              Asian   0.0100188 0.103210  0.0970728 0.922669
        Partyid                         Republican  -0.1397397 0.092869 -1.5047041 0.132400
        Partyid                           Democrat   0.1568092 0.068036  2.3047920 0.021178 *
---
Number of Obs. = 1500
Number of Respondents = 75
---
Signif. codes: 0 '***' 0.001 '**' 0.01 '*' 0.05
```

Table 3.3 - *Conditional effects of a candidate's attributes on respondents' self-assessed interest in politics (Moderately interested)*

```
-------------------------------------------------------------
Conditional AMCE's (Resppolint = Moderately interested):
-------------------------------------------------------------
      Attribute                      Level     Estimate Std. Err    z value   Pr(>|z|)
Pol_prominence    Locally Renowned Party Member  0.1583840 0.060558   2.615427 8.9116e-03  **
Pol_prominence  Statewide Renowned Party Member  0.2098561 0.073445   2.857326 4.2723e-03  **
Pol_prominence Nationally Renowned Party Member  0.1071141 0.069126   1.549549 1.2125e-01
Pub_prominence              2.400 Twitter followers -0.0328636 0.075189 -0.437082 6.6205e-01
Pub_prominence             23.700 Twitter followers -0.0165045 0.080426 -0.205213 8.3741e-01
Pub_prominence            315.000 Twitter followers  0.0485816 0.081260  0.597850 5.4994e-01
Pub_prominence        1.3 Million Twitter followers  0.0628314 0.082228  0.764116 4.4480e-01
        Income             Annual income $54.000 -0.0687013 0.106830 -0.643088 5.2017e-01
        Income             Annual income $75.000 -0.0423806 0.089446 -0.473811 6.3563e-01
        Income             Annual income $92.000 -0.1504425 0.106850 -1.407974 1.5914e-01
        Income            Annual income $140.000 -0.0928908 0.082903 -1.120476 2.6251e-01
        Income            Annual income $360.000 -0.0654248 0.079676 -0.821132 4.1157e-01
        Income            Annual income $840.000 -0.0942980 0.093770 -1.005631 3.1459e-01
           Age                      38 years old -0.0378451 0.073275 -0.516480 6.0552e-01
           Age                      45 years old -0.0855220 0.082802 -1.032848 3.0168e-01
           Age                      52 years old -0.1009890 0.067035 -1.506512 1.3194e-01
           Age                      59 years old -0.1005990 0.079834 -1.260107 2.0763e-01
           Age                      66 years old -0.2542427 0.059219 -4.293274 1.7606e-05 ***
           Age                      73 years old -0.1555817 0.080806 -1.925367 5.4183e-02
    Occupation                  Military Officer  0.0025767 0.105446  0.024437 9.8050e-01
    Occupation                           Teacher  0.0388837 0.086449  0.449789 6.5286e-01
    Occupation                            Farmer -0.0220325 0.102801 -0.214321 8.3030e-01
    Occupation                    Business owner -0.0485780 0.087427 -0.555639 5.7846e-01
    Occupation                           Athlete -0.0848714 0.085621 -0.991246 3.2157e-01
    Occupation                             Actor -0.1225990 0.107502 -1.140432 2.5411e-01
    Occupation                            Banker -0.1876416 0.128844 -1.456350 1.4530e-01
    Occupation                        Journalist -0.1524348 0.098803 -1.542816 1.2288e-01
    Occupation                      Union Leader -0.0737937 0.110070 -0.670422 5.0259e-01
        Gender                              Male -0.0432064 0.050250 -0.859829 3.8988e-01
    Incumbency                        Challenger -0.0649619 0.041644 -1.559951 1.1877e-01
     Ethnicity                  African American -0.0015187 0.096675 -0.015709 9.8747e-01
     Ethnicity                  Hispanic/ Latino -0.0065287 0.076217 -0.085660 9.3174e-01
     Ethnicity                   Native American -0.0690877 0.076230 -0.906302 3.6478e-01
     Ethnicity                             Asian -0.0362031 0.087139 -0.415462 6.7780e-01
       Partyid                        Republican -0.1276472 0.071565 -1.783666 7.4478e-02
       Partyid                          Democrat  0.1251167 0.067388  1.856673 6.3358e-02
---
Number of Obs. = 1500
Number of Respondents = 75
---
Signif. codes: 0 '***' 0.001 '**' 0.01 '*' 0.05
```

Table 3.4 - *Conditional effects of a candidate's attributes on respondents' self-assessed interest in politics (Rather interested)*

```
-------------------------------------------------------------
Conditional AMCE's (Resppolint = Rather interested):
-------------------------------------------------------------
       Attribute                         Level    Estimate  Std. Err    z value   Pr(>|z|)
 Pol_prominence     Locally Renowned Party Member  0.077495  0.080744   0.959768  0.337172
 Pol_prominence   Statewide Renowned Party Member -0.035895  0.066838  -0.537044  0.591237
 Pol_prominence  Nationally Renowned Party Member  0.131723  0.065902   1.998761  0.045634 *
 Pub_prominence              2.400 Twitter followers -0.040758  0.068804  -0.592370  0.553603
 Pub_prominence             23.700 Twitter followers -0.081178  0.055925  -1.451548  0.146627
 Pub_prominence            315.000 Twitter followers -0.140424  0.067918  -2.067558  0.038682 *
 Pub_prominence         1.3 Million Twitter followers -0.035999  0.057246  -0.628843  0.529452
         Income           Annual income $54.000 -0.091701  0.087999  -1.042080  0.297375
         Income           Annual income $75.000 -0.068730  0.088767  -0.774274  0.438769
         Income           Annual income $92.000 -0.047801  0.112875  -0.423490  0.671938
         Income          Annual income $140.000 -0.121657  0.144604  -0.841309  0.400175
         Income          Annual income $360.000 -0.145271  0.148345  -0.979280  0.327442
         Income          Annual income $840.000 -0.267941  0.122253  -2.191695  0.028402 *
            Age                    38 years old -0.049463  0.066569  -0.743033  0.457462
            Age                    45 years old  0.043946  0.091440   0.480603  0.630799
            Age                    52 years old -0.124450  0.088631  -1.404135  0.160279
            Age                    59 years old -0.161987  0.097057  -1.668995  0.095118
            Age                    66 years old -0.012577  0.104332  -0.120544  0.904053
            Age                    73 years old -0.107947  0.102883  -1.049218  0.294078
     Occupation                Military Officer -0.064673  0.096215  -0.672175  0.501472
     Occupation                         Teacher  0.116703  0.110872   1.052590  0.292529
     Occupation                          Farmer  0.012357  0.128368   0.096264  0.923311
     Occupation                  Business owner  0.047113  0.117408   0.401281  0.688213
     Occupation                         Athlete -0.164321  0.138355  -1.187674  0.234962
     Occupation                           Actor -0.084295  0.131122  -0.642871  0.520308
     Occupation                          Banker  0.084474  0.117514   0.718836  0.472242
     Occupation                       Journalist  0.042042  0.111130   0.378318  0.705195
     Occupation                     Union Leader  0.155959  0.131501   1.185991  0.235626
         Gender                            Male  0.039964  0.045999   0.868797  0.384958
     Incumbency                       Challenger  0.010171  0.043751   0.232482  0.816164
      Ethnicity                African American -0.138500  0.079811  -1.735350  0.082679
      Ethnicity                 Hispanic/ Latino  0.050768  0.074348   0.682846  0.494704
      Ethnicity                  Native American  0.048207  0.075704   0.636787  0.524264
      Ethnicity                           Asian  0.092736  0.062200   1.490930  0.135980
        Partyid                       Republican -0.152246  0.085919  -1.771960  0.076401
        Partyid                         Democrat  0.083133  0.079134   1.050538  0.293471
---
Number of Obs. = 1500
Number of Respondents = 75
---
Signif. codes: 0 '***' 0.001 '**' 0.01 '*' 0.05
```

Table 3.5 - *Conditional effects of a candidate's attributes on respondents' self-assessed interest in politics (Very interested)*

Table 4.1 - *Conditional effects of a candidate's attributes on respondents' level of education (Some high school, not completed).*

```
----------------------------------------------------------
Conditional AMCE's (Respeduc = Highschool, no diploma):
----------------------------------------------------------
      Attribute                         Level   Estimate  Std. Err      z value     Pr(>|z|)
  Pol_prominence    Locally Renowned Party Member  0.7585441 0.0273138   27.77143 9.6030e-170 ***
  Pol_prominence  Statewide Renowned Party Member -0.1135063 0.0202144   -5.61511  1.9644e-08 ***
  Pol_prominence Nationally Renowned Party Member  1.0843424 0.0638124   16.99265  9.3086e-65 ***
  Pub_prominence            2.400 Twitter followers  0.5480743 0.0668829    8.19453  2.5156e-16 ***
  Pub_prominence           23.700 Twitter followers  0.6845040 0.0999931    6.84552  7.6201e-12 ***
  Pub_prominence          315.000 Twitter followers  0.6383023 0.0589079   10.83560  2.3342e-27 ***
  Pub_prominence      1.3 Million Twitter followers  0.8286061 0.0335105   24.72673 5.5182e-135 ***
          Income           Annual income $54.000 -0.0499690 0.0589300   -0.84794  3.9647e-01
          Income           Annual income $75.000  0.1377032 0.0980042    1.40508  1.6000e-01
          Income           Annual income $92.000  0.0360784 0.1091946    0.33040  7.4109e-01
          Income          Annual income $140.000  0.6475456 0.1337816    4.84032  1.2963e-06 ***
          Income          Annual income $360.000 -0.1599027 0.0505778   -3.16152  1.5695e-03 **
          Income          Annual income $840.000  0.7174624 0.0257917   27.81760 2.6569e-170 ***
             Age                     38 years old -0.0277107 0.0076011   -3.64562  2.6675e-04 ***
             Age                     45 years old -0.1411238 0.0215129   -6.55995  5.3824e-11 ***
             Age                     52 years old -0.3136270 0.0552677   -5.67468  1.3894e-08 ***
             Age                     59 years old -0.2415869 0.0026821  -90.07229  0.0000e+00 ***
             Age                     66 years old -0.6103570 0.0812235   -7.51454  5.7113e-14 ***
             Age                     73 years old -0.0757536 0.0565835   -1.33879  1.8064e-01
      Occupation                  Military Officer -1.3068903 0.0294615  -44.35920  0.0000e+00 ***
      Occupation                           Teacher -1.5904632 0.0838595  -18.96580  3.2700e-80 ***
      Occupation                            Farmer -1.2942030 0.0484782  -26.69661 5.1527e-157 ***
      Occupation                    Business owner -1.1628119 0.0894706  -12.99657  1.2795e-38 ***
      Occupation                           Athlete -2.4098315 0.0872041  -27.63438 4.3004e-168 ***
      Occupation                             Actor -1.6965581 0.0550662  -30.80942 1.9596e-208 ***
      Occupation                            Banker -2.0333851 0.1032199  -19.69954  2.1757e-86 ***
      Occupation                        Journalist -1.6770146 0.0066828 -250.94672  0.0000e+00 ***
      Occupation                      Union Leader -1.5427431 0.0161216  -95.69437  0.0000e+00 ***
          Gender                              Male  0.0888650 0.0125895    7.05864  1.6814e-12 ***
      Incumbency                        Challenger -0.5589187 0.0293821  -19.02239  1.1128e-80 ***
       Ethnicity                  African American  0.0722783 0.0125546    5.75713  8.5557e-09 ***
       Ethnicity                   Hispanic/ Latino  0.4383456 0.0425475   10.30251  6.8648e-25 ***
       Ethnicity                   Native American  0.7607349 0.0753996   10.08938  6.1558e-24 ***
       Ethnicity                             Asian -0.0237468 0.0513908   -0.46208  6.4402e-01
         Partyid                        Republican  0.2694240 0.0129716   20.77028  8.0391e-96 ***
         Partyid                          Democrat  0.0088658 0.0764250    0.11601  9.0765e-01
---
Number of Obs. = 1500
Number of Respondents = 75
---
Signif. codes: 0 '***' 0.001 '**' 0.01 '*' 0.05
```

Table 4.2 - *Conditional effects of a candidate's attributes on respondents' level of education (High school diploma).*

```
-------------------------------------------------------------
Conditional AMCE's (Respeduc = Highschool diploma or equivalent):
-------------------------------------------------------------
      Attribute                              Level   Estimate Std. Err    z value   Pr(>|z|)
  Pol_prominence     Locally Renowned Party Member  0.0170555 0.050006   0.341069 7.3305e-01
  Pol_prominence   Statewide Renowned Party Member -0.0067021 0.058855  -0.113875 9.0934e-01
  Pol_prominence  Nationally Renowned Party Member  0.0032258 0.069056   0.046712 9.6274e-01
  Pub_prominence              2.400 Twitter followers -0.1016977 0.114594 -0.887461 3.7483e-01
  Pub_prominence             23.700 Twitter followers -0.0449909 0.151018 -0.297918 7.6577e-01
  Pub_prominence            315.000 Twitter followers -0.0453621 0.155385 -0.291933 7.7034e-01
  Pub_prominence          1.3 Million Twitter followers -0.0113247 0.120701 -0.093824 9.2525e-01
          Income              Annual income $54.000  0.2501327 0.087563   2.856591 4.2822e-03 **
          Income              Annual income $75.000  0.2530348 0.102350   2.472242 1.3427e-02 *
          Income              Annual income $92.000  0.0980027 0.102967   0.951788 3.4120e-01
          Income             Annual income $140.000 -0.0088500 0.077607  -0.114036 9.0921e-01
          Income             Annual income $360.000 -0.0152842 0.128908  -0.118566 9.0562e-01
          Income             Annual income $840.000  0.0427834 0.091615   0.466992 6.4051e-01
             Age                       38 years old -0.2787787 0.143283  -1.945654 5.1696e-02
             Age                       45 years old -0.3533120 0.072593  -4.867010 1.1330e-06 ***
             Age                       52 years old -0.4103281 0.121972  -3.364131 7.6785e-04 ***
             Age                       59 years old -0.5074533 0.187136  -2.711687 6.6942e-03 **
             Age                       66 years old -0.5228028 0.110230  -4.742841 2.1074e-06 ***
             Age                       73 years old -0.2570233 0.125741  -2.044074 4.0946e-02 *
      Occupation                    Military Officer -0.1436752 0.136565 -1.052064 2.9277e-01
      Occupation                            Teacher  0.0819262 0.151451  0.540941 5.8855e-01
      Occupation                             Farmer -0.1447513 0.130181 -1.111926 2.6617e-01
      Occupation                      Business owner  0.0408812 0.051531  0.793330 4.2759e-01
      Occupation                            Athlete -0.0954626 0.140845 -0.677783 4.9791e-01
      Occupation                              Actor -0.2583176 0.156411 -1.651525 9.8631e-02
      Occupation                             Banker  0.1904504 0.164579  1.157199 2.4719e-01
      Occupation                          Journalist  0.0420031 0.170968  0.245677 8.0593e-01
      Occupation                        Union Leader  0.0394330 0.179312  0.219913 8.2594e-01
          Gender                              Male  0.0547772 0.055860  0.980622 3.2678e-01
       Incumbency                        Challenger -0.0985762 0.067307 -1.464567 1.4304e-01
        Ethnicity                   African American  0.0452658 0.134108  0.337532 7.3572e-01
        Ethnicity                   Hispanic/ Latino -0.0057583 0.123058 -0.046793 9.6268e-01
        Ethnicity                    Native American -0.0318309 0.115115 -0.276514 7.8215e-01
        Ethnicity                              Asian -0.0220217 0.108814 -0.202379 8.3962e-01
          Partyid                          Republican -0.1228314 0.113877 -1.078636 2.8075e-01
          Partyid                            Democrat  0.0606567 0.058977  1.028477 3.0373e-01
---
Number of Obs. = 1500
Number of Respondents = 75
---
Signif. codes: 0 '***' 0.001 '**' 0.01 '*' 0.05
```

Table 4.3 - *Conditional effects of a candidate's attributes on respondents' level of education (Some college or university, not completed).*

```
-------------------------------------------------------------
Conditional AMCE's (Respeduc = Some college or university studies, not completed):
-------------------------------------------------------------
         Attribute                             Level    Estimate Std. Err    z value    Pr(>|z|)
    Pol_prominence    Locally Renowned Party Member    0.1548794 0.084800   1.826412 6.7788e-02
    Pol_prominence  Statewide Renowned Party Member    0.1661852 0.104394   1.591899 1.1141e-01
    Pol_prominence Nationally Renowned Party Member    0.1826891 0.079376   2.301555 2.1360e-02  *
    Pub_prominence              2.400 Twitter followers -0.0990852 0.092339 -1.073057 2.8325e-01
    Pub_prominence             23.700 Twitter followers -0.1364759 0.083141 -1.641508 1.0069e-01
    Pub_prominence            315.000 Twitter followers -0.0035299 0.082882 -0.042589 9.6603e-01
    Pub_prominence        1.3 Million Twitter followers -0.0249314 0.088395 -0.282047 7.7791e-01
            Income           Annual income $54.000      0.0289840 0.115424   0.251109 8.0173e-01
            Income           Annual income $75.000     -0.0964364 0.061144  -1.577207 1.1475e-01
            Income           Annual income $92.000     -0.2538430 0.133344  -1.903665 5.6954e-02
            Income          Annual income $140.000     -0.2769711 0.066221  -4.182503 2.8832e-05 ***
            Income          Annual income $360.000     -0.1634807 0.117751  -1.388361 1.6503e-01
            Income          Annual income $840.000     -0.3068197 0.118403  -2.591327 9.5607e-03  **
               Age                       38 years old  -0.1292844 0.104682  -1.235023 2.1682e-01
               Age                       45 years old  -0.1183770 0.097282  -1.216845 2.2366e-01
               Age                       52 years old  -0.1531925 0.100964  -1.517296 1.2919e-01
               Age                       59 years old  -0.0973413 0.100533  -0.968253 3.3292e-01
               Age                       66 years old  -0.2194776 0.088669  -2.475235 1.3315e-02  *
               Age                       73 years old  -0.1166788 0.090912  -1.283420 1.9935e-01
        Occupation                   Military Officer   0.0194332 0.151524   0.128251 8.9795e-01
        Occupation                            Teacher   0.0706676 0.147304   0.479739 6.3141e-01
        Occupation                             Farmer   0.0361596 0.166014   0.217811 8.2758e-01
        Occupation                     Business owner  -0.0133592 0.135852  -0.098337 9.2167e-01
        Occupation                            Athlete  -0.1750452 0.166009  -1.054434 2.9168e-01
        Occupation                              Actor  -0.2475638 0.140584  -1.760971 7.8243e-02
        Occupation                             Banker  -0.1176594 0.137944  -0.852950 3.9369e-01
        Occupation                         Journalist  -0.1418563 0.140744  -1.007905 3.1350e-01
        Occupation                       Union Leader   0.0079986 0.164263   0.048694 9.6116e-01
            Gender                               Male   0.0099547 0.063465   0.156855 8.7536e-01
        Incumbency                         Challenger   0.0155273 0.049316   0.314854 7.5287e-01
         Ethnicity                   African American  -0.1729142 0.098283  -1.759347 7.8519e-02
         Ethnicity                   Hispanic/ Latino  -0.0788259 0.070514  -1.117873 2.6362e-01
         Ethnicity                    Native American  -0.1060651 0.087956  -1.205887 2.2786e-01
         Ethnicity                              Asian  -0.0948309 0.083688  -1.133152 2.5715e-01
           Partyid                         Republican  -0.2165348 0.094378  -2.294332 2.1771e-02  *
           Partyid                           Democrat   0.1831745 0.102467   1.787645 7.3833e-02
---
Number of Obs. = 1500
Number of Respondents = 75
---
Signif. codes: 0 '***' 0.001 '**' 0.01 '*' 0.05
```

Table 4.4 - *Conditional effects of a candidate's attributes on respondents' level of education (College or university, completed).*

```
---------------------------------------------------------------
Conditional AMCE's (Respeduc = College or university studies, completed):
---------------------------------------------------------------
        Attribute                            Level    Estimate  Std. Err    z value  Pr(>|z|)
    Pol_prominence      Locally Renowned Party Member  0.04310419 0.052513   0.820832 0.411742
    Pol_prominence    Statewide Renowned Party Member  0.02089242 0.052949   0.394574 0.693157
    Pol_prominence   Nationally Renowned Party Member  0.10035715 0.056893   1.763970 0.077737
    Pub_prominence                2.400 Twitter followers  0.00977212 0.057497   0.169959 0.865042
    Pub_prominence               23.700 Twitter followers -0.00096752 0.061438  -0.015748 0.987436
    Pub_prominence              315.000 Twitter followers -0.00337880 0.067089  -0.050363 0.959833
    Pub_prominence          1.3 Million Twitter followers  0.06388954 0.058846   1.085701 0.277611
            Income              Annual income $54.000  -0.05122028 0.071408  -0.717289 0.473196
            Income              Annual income $75.000  -0.02742468 0.067782  -0.404602 0.685770
            Income              Annual income $92.000  -0.03085220 0.080705  -0.382285 0.702250
            Income             Annual income $140.000  -0.08578578 0.080876  -1.060713 0.288820
            Income             Annual income $360.000  -0.06292051 0.087387  -0.720019 0.471513
            Income             Annual income $840.000  -0.15433323 0.083603  -1.846026 0.064888
               Age                       38 years old   0.01375520 0.051982   0.264613 0.791308
               Age                       45 years old   0.12310228 0.058372   2.108936 0.034950 *
               Age                       52 years old  -0.06521082 0.067714  -0.963032 0.335531
               Age                       59 years old  -0.08377005 0.061807  -1.355339 0.175310
               Age                       66 years old  -0.08602176 0.063042  -1.364507 0.172408
               Age                       73 years old  -0.07989039 0.078225  -1.021291 0.307116
        Occupation                   Military Officer   0.00693671 0.085725   0.080918 0.935507
        Occupation                            Teacher   0.05349582 0.071420   0.749033 0.453837
        Occupation                             Farmer   0.06045709 0.083755   0.721835 0.470396
        Occupation                     Business owner  -0.09522914 0.069582  -1.368594 0.171126
        Occupation                            Athlete  -0.01209448 0.082812  -0.146047 0.883884
        Occupation                              Actor   0.01923036 0.080330   0.239392 0.810802
        Occupation                             Banker   0.04545519 0.095184   0.477553 0.632969
        Occupation                         Journalist   0.05590233 0.085318   0.655226 0.512322
        Occupation                       Union Leader   0.09099113 0.082573   1.101951 0.270483
            Gender                               Male  -0.01859769 0.034510  -0.538909 0.589950
        Incumbency                          Challenger -0.04568515 0.035900  -1.272555 0.203176
         Ethnicity                   African American  -0.03142026 0.070612  -0.444972 0.656340
         Ethnicity                     Hispanic/Latino  0.01688189 0.056243   0.300162 0.764054
         Ethnicity                     Native American  0.01541700 0.059833   0.257667 0.796664
         Ethnicity                               Asian  0.06784411 0.061679   1.099948 0.271355
           Partyid                          Republican -0.08246931 0.063201  -1.304863 0.191939
           Partyid                            Democrat  0.11371005 0.053414   2.128852 0.033266 *
---
Number of Obs. = 1500
Number of Respondents = 75
---
Signif. codes:  0 '***' 0.001 '**' 0.01 '*' 0.05
```

Table 4.5 - *Conditional effects of a candidate's attributes on respondents' level of education (Graduate studies).*

```
----------------------------------------------------------
Conditional AMCE's (Respeduc = Graduate studies):
----------------------------------------------------------
       Attribute                           Level   Estimate  Std. Err    z value    Pr(>|z|)
   Pol_prominence     Locally Renowned Party Member   0.189591  0.097704   1.94046  5.2324e-02   .
   Pol_prominence   Statewide Renowned Party Member   0.177645  0.093472   1.90052  5.7365e-02   .
   Pol_prominence  Nationally Renowned Party Member   0.259774  0.115369   2.25167  2.4343e-02   *
   Pub_prominence               2.400 Twitter followers  -0.239397  0.181300  -1.32045  1.8669e-01
   Pub_prominence              23.700 Twitter followers   0.205991  0.097005   2.12351  3.3711e-02   *
   Pub_prominence             315.000 Twitter followers   0.051190  0.120011   0.42654  6.6971e-01
   Pub_prominence         1.3 Million Twitter followers   0.187473  0.175744   1.06674  2.8609e-01
           Income              Annual income $54.000    0.170733  0.103854   1.64396  1.0018e-01
           Income              Annual income $75.000   -0.227015  0.173767  -1.30643  1.9141e-01
           Income              Annual income $92.000    0.256086  0.164382   1.55787  1.1926e-01
           Income             Annual income $140.000   -0.231124  0.173918  -1.32893  1.8387e-01
           Income             Annual income $360.000   -0.189060  0.172660  -1.09498  2.7352e-01
           Income             Annual income $840.000   -0.114963  0.113802  -1.01020  3.1240e-01
              Age                         38 years old   0.247136  0.169637   1.45685  1.4516e-01
              Age                         45 years old   0.526005  0.085274   6.16842  6.8976e-10  ***
              Age                         52 years old   0.057396  0.120289   0.47715  6.3326e-01
              Age                         59 years old   0.105717  0.112684   0.93817  3.4816e-01
              Age                         66 years old   0.057006  0.105129   0.54225  5.8765e-01
              Age                         73 years old   0.134080  0.155541   0.86203  3.8867e-01
       Occupation                     Military Officer  -0.193299  0.231962  -0.83332  4.0466e-01
       Occupation                              Teacher   0.089057  0.164780   0.54046  5.8888e-01
       Occupation                               Farmer  -0.568551  0.141235  -4.02557  5.6836e-05  ***
       Occupation                       Business owner  -0.199792  0.116308  -1.71778  8.5837e-02   .
       Occupation                              Athlete  -0.132105  0.163612  -0.80743  4.1942e-01
       Occupation                                Actor  -0.275929  0.109704  -2.51521  1.1896e-02   *
       Occupation                               Banker  -0.880067  0.159652  -5.51239  3.5399e-08  ***
       Occupation                           Journalist  -0.666667  0.295534  -2.25581  2.4083e-02   *
       Occupation                         Union Leader  -0.446868  0.146387  -3.05266  2.2683e-03  **
           Gender                                 Male  -0.138046  0.047998  -2.87606  4.0267e-03  **
        Incumbency                           Challenger   0.213755  0.169401   1.26183  2.0701e-01
         Ethnicity                     African American   0.192677  0.112149   1.71806  8.5787e-02   .
         Ethnicity                      Hispanic/ Latino  -0.151541  0.074940  -2.02216  4.3160e-02   *
         Ethnicity                      Native American   0.397887  0.116074   3.42787  6.0835e-04  ***
         Ethnicity                                Asian   0.454497  0.068229   6.66130  2.7141e-11  ***
           Partyid                           Republican  -0.315110  0.139107  -2.26524  2.3498e-02   *
           Partyid                             Democrat   0.198466  0.155365   1.27742  2.0145e-01
---
Number of Obs. = 1500
Number of Respondents = 75
---
Signif. codes: 0 '***' 0.001 '**' 0.01 '*' 0.05
```